\documentclass[conference, a4paper]{IEEEtran}
\PassOptionsToPackage{hyphens}{url}\usepackage[hidelinks]{hyperref}
\ifCLASSINFOpdf
  \usepackage[pdftex]{graphicx}
  \graphicspath{{figs/pdf/}{figs/jpeg/}}
\else
\fi
\def\BibTeX{{\rm B\kern-.05em{\sc i\kern-.025em b}\kern-.08em
    T\kern-.1667em\lower.7ex\hbox{E}\kern-.125emX}}
\usepackage{cite}
\usepackage[cmex10]{amsmath}
\usepackage{siunitx}
\usepackage{subfigure}
\usepackage{multirow}
\begin{document}
\title{WIP: Achieving Self-Interference-Free Operation on SDR Platform with Critical TDD Turnaround Time}

\author{\IEEEauthorblockN{Thijs Havinga, Xianjun Jiao, Muhammad Aslam, Wei Liu and Ingrid Moerman}
\IEEEauthorblockA{\textit{IDLab, Ghent University - imec, Belgium} \\
Email: \{Thijs.Havinga, Xianjun.Jiao, Muhammad.Aslam, Wei.Liu, Ingrid.Moerman\}@UGent.be}
}

\maketitle

\begin{abstract}
Software Defined Radio (SDR) platforms are valuable for research and development activities or high-end systems that demand real-time adaptable wireless protocols. While low latency can be achieved using the dedicated digital processing unit of a state-of-the-art SDR platform, its Radio Frequency (RF) front-end often poses a limitation in terms of turnaround time (TT), the time needed for switching from the receiving to the transmitting mode (or vice versa). Zero Intermediate Frequency (ZIF) transceivers are favorable for SDR, but suffer from self-interference even if the device is not currently transmitting. The strict MAC-layer requirements of Time Division Duplex (TDD) protocols like Wi-Fi cannot be achieved using configurable ZIF transceivers without having to compromise receiver sensitivity. Using a novel approach, we show that the TT using the AD9361 RF front-end can be as low as \SI{640}{ns}, while the self-interference is at the same level as achieved by the conventional TDD mode, which has a TT of at least \SI{55}{\micro \second}. As compared to Frequency Division Duplex (FDD) mode, a decrease of receiver noise floor by about \SI{13}{dB} in the \SI{2.4}{GHz} band and by about \SI{4.5}{dB} in the \SI{5}{GHz} band is achieved. 
\end{abstract}


\section{Introduction}
Software Defined Radio (SDR) platforms enable researchers to experiment and prototype innovative wireless communication solutions more freely as compared to Commercial Off-The-Shelf (COTS) chips thanks to their flexibility and openness. In addition, the configurable nature of SDR makes it also a viable platform for high-end solutions in environments with dynamic characteristics or requirements.

An SDR generally consists of two parts. The first, a processing unit, performs the digital signal processing operations, such as modulating and demodulating. Next, an SDR has a configurable Radio Frequency (RF) front-end, which consists amongst others of a Digital-to-Analog Converter (DAC) and Analog-to-Digital Converter (ADC), an amplifier, mixer and several filters. The processing unit can be a separate host computer by using its Central Processing Unit (CPU). Another option is to use a Field Programmable Gate Array (FPGA), in which digital hardware is configured by how the designer programs it.

Currently, one of the downsides of SDR using a host computer is the latency induced by the data travelling between the radio front-end and the processing unit. For example, according to \cite{Jiao18}, the latency between a host computer and a Universal Software Radio Peripheral (USRP) version X310 using a PCIe link is \SI{79}{\micro \second}. For applications that require low latency, an SDR based on a System-on-Chip (SoC), where the connection between CPU, FPGA and RF front-end is fast, is more suitable. For example, the connection between CPU and FPGA can be as low as \SI{1.435}{\micro \second} \cite{Jiao18}. 

For the implementation of Time Division Duplex (TDD) systems, the SDR needs to switch from the receiving (Rx) to the transmitting (Tx) mode (and vice versa). The time this takes is called the turnaround time (TT). The design of a Medium Access Control (MAC) protocol is limited by the TT. For instance in the Wi-Fi standard, an acknowledgement frame as a response to a successfully received frame should arrive at the transmitter within a minimum time, called Short InterFrame Space (SIFS). For the standards IEEE 802.11a/g/n, the SIFS is \SI{10}{\micro \second} when operating in the \SI{2.4}{GHz} band and \SI{16}{\micro \second} in the \SI{5}{GHz} bands \cite{IEEE16}. In the 5G standard \cite{ETSI}, a guard period of \SI{17.84}{\micro \second} is specified for a downlink to uplink pattern when using a subcarrier spacing of \SI{120}{kHz}. 

The TT of protocol-specific chips is in the order of \SI{1}{\micro \second} \cite{maximintegrated}, whereas configurable RF front-ends have significantly higher TT. In this paper, we consider Zero-Intermediate Frequency (ZIF) transceivers, more specifically the AD9361 front-end \cite{AD9361}, which is widely used in the SDR research community. ZIF transceivers use a local oscillator (LO) that produces a frequency equal to or close to the carrier frequency. In this way, less components are needed as compared to transceivers using an intermediate frequency, which makes it more suited for implementation on a chip.
The lowest supported TT of the AD9361 is \SI{18}{\micro \second} according to the specification \cite{tddswitchingtime}, which is too high for many MAC protocols. Operation in Frequency Division Duplex (FDD) mode, where both the Rx and Tx chain are active, avoids this TT, but suffers from receiver sensitivity degradation due to Tx LO leakage when operating at the same frequency. This leakage is present even if the system is not currently transmitting. It also cannot be eliminated by conventional filtering, since the noise exists in the same frequency range as the desired signal. 

In this paper, we propose a method for self-interference-free operation of a ZIF transceiver while achieving a low turnaround time. For this, we consider the implementation of Wi-Fi using the \textit{openwifi} platform \cite{jiao2020openwifi}, which is suited for SDRs based on a SoC. First, we discuss the related work on this topic. Next, we explore different options before presenting our proposed solution. Then, we show and evaluate the results of measurements on a real hardware platform, after which we conclude the paper. 

\section{Related work}
The authors of \cite{Tripathi19} propose to use a digital intermediate frequency shift to avoid LO leakage. They shift the baseband signal by \SI{5}{MHz}, such that it falls outside of their \SI{5}{MHz} bandwidth. The in-band error between a transmitted and captured signal was measured using the AD9361 RF front-end when transmitter and receiver were connected via a cable. In order to eliminate the I/Q imbalance image that is created by the shift, a baseband filter was needed. After this filter, the normalized mean square error was reduced by about \SI{10}{dB}. Since this method cannot completely eliminate the self-interference and leads to additional latency and resources when filtering the spectral images, its performance gain is only limited. 

In \cite{Muhammad18}, a method to achieve zero TT by leaving both the Tx and Rx chain of the RF front-end on is presented. The switching between Tx and Rx is done in the digital baseband domain, which has separate processing units for Tx and Rx. Using a set-up with separate antennas placed exactly orthogonal, there was still self-interference of \SI{3.2}{dB}.

Implementation of Wi-Fi meeting the SIFS requirement using a ZIF transceiver has already been done in \cite{Wu18Tick}. The authors use the AD9371, which has similar TT as the AD9361, to implement IEEE 802.11ac and full-duplex 802.11a/g. To meet the SIFS, their system consists of a separate Rx and Tx setup, which do not switch states. Such a setup does not represent a realistic system and will therefore not be suited for practical research or real-life deployment.

In \cite{Seijo21}, 2-3\si{\micro \second} Rx-Tx switching delay is assumed for the calculation of a minimum interframe space for their custom MAC protocol. The authors show with their hardware platform using the AD9361 radio chip that an interframe space of \SI{12}{\micro \second} can be achieved. For this set-up, it is not clear whether Rx-Tx switching is done, nor do they mention the receiver sensitivity.

In order to realize full-duplex systems on SDR platforms, the works \cite{Wu16GRT}, \cite{Anttila21} and \cite{Amjad20} present digital self-interference mitigation techniques. Such techniques, however, cannot eliminate the self-interference to the same extent as compared to switching between Rx and Tx due to the unpredictable noise coming from the LO. Furthermore, it comes at the cost of additional computation time or hardware resources.

\section{Proposed Solution} 
\label{sec:solution}
Here we discuss possible solutions to eliminate self-interference using the widely used AD9361 and a SoC, while still maintaining low latency. The idea behind these solutions can in general be applied to all ZIF transceivers.

\begin{figure}
    \centering
    \includegraphics[width=0.8\linewidth]{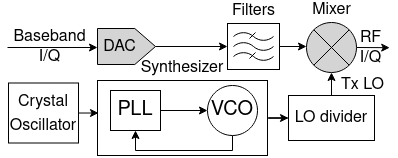}
    \caption{Simplified diagram of the AD9361 Tx path.}
    \label{fig:AD9361Tx}
    \vspace{-0.2cm}
\end{figure}

A simplified diagram of the AD9361 Tx path is shown in Figure \ref{fig:AD9361Tx}. 
The AD9361 can operate in Frequency Division Duplex (FDD) or Time Division Duplex (TDD) mode. In FDD mode, both Rx and Tx synthesizers remain on at all times, whereas in TDD operation these are powered up alternately. Furthermore, there are some options to tweak the operations, which are listed hereafter \cite{tddswitchingtime}: 
\begin{enumerate}
    \item \textit{Standard Enable State Machine (ENSM) TDD Mode}: In this mode, Voltage-Controlled Oscillator (VCO) calibrations take place when the internal state machine changes from Rx to Tx state. The calibration time depends on the reference clock frequency, but takes at least \SI{37}{\micro \second}. After these calibrations, the Rx or Tx Phase-Locked Loop (PLL) needs to lock, which takes about \SI{15}{\micro \second}. Furthermore, when changing to Tx, the DAC needs to power up, which takes about \SI{18}{\micro \second}. Also, the Tx and Rx data paths need to be flushed, which takes 384 ADC clock cycles. E.g., for a typical ADC clock frequency of \SI{160}{MHz}, this boils down to \SI{2.4}{\micro \second}. The latter two operations can be done in parallel to the PLL lock, so this method has a TT of at least \SI{55}{\micro \second}.
    \item \textit{Standard TDD Mode}: This mode is the same as the one mentioned above, except that calibrations are disabled. When the LO frequency does not change from frame to frame, it is not necessary to re-calibrate every time. Thus, the TT using this method is dominated by the DAC power up time of \SI{18}{\micro \second}.
    \item \textit{Standard TDD Mode/Dual Synth}: 
    In this case, no calibrations take place and the PLLs do not have to be locked, since these are on at all times. However, the DAC still needs to power up and flushing needs to be done, resulting in at least \SI{18}{\micro \second} delay. 
    \item \textit{FDD Independent Mode}: In this mode, both the Rx and Tx data paths are active, but can be independently controlled. No calibrations or flushing takes place, but the Tx DAC still powers up, resulting in \SI{18}{\micro \second} delay. 
\end{enumerate}
None of these methods can comply to strict MAC requirements like SIFS, thus we are committed to normal FDD mode. The AD9361 already provides RF DC offset and quadrature tracking calibration to mitigate phase and gain error, but these do not suffice to fully eliminate the self-interference, especially when antennas are attached to the Tx and Rx ports. Therefore, we propose the following method to further suppress self-interference in FDD mode. Instead of powering down the synthesizers in the AD9361, there is an option to power down only the LO divider, such that the signal does not reach the mixer \cite{AD9361register}. Powering on the LO divider takes about \SI{160}{ns} \cite{ezanalog}. In FDD mode, both chains remain on completely, so there is no additional powering up or calibration needed. 

Turning the LO on and off can only be done via an SPI write to the AD9361. A CPU connected to the AD9361 can issue such an SPI write, but this does not ensure real-time operation due to its unpredictable response time. Therefore, we implement SPI functionality in the FPGA. We use the maximum SPI frequency of the AD9361 (i.e. \SI{50}{MHz}) and 24 bits are needed for the instruction to turn the Tx LO on or off, resulting in a duration of \SI{480}{ns}. Thus, the TT is reduced to \SI{640}{ns}, which is well within the SIFS requirement of Wi-Fi.

\section{Experimental results and evaluation}
In this section, two aspects of the proposed solution are examined. First, TT is quantified by measuring the Tx power over time. Secondly, the impact on receiver sensitivity is assessed by performing receiver noise floor tests. The hardware used is the Xilinx Zynq UltraScale+ MPSoC ZCU102, with an AD-FMCOMMS2-EBZ front-end (containing the AD9361) connected via an FPGA Mezzanine Card (FMC) interface. The SPI module is implemented using Verilog and its outputs are connected to the SPI interface of the AD9361. The module uses only 33 lookup tables and 28 registers and can be found online \cite{openwifi} in \textit{openwifi-hw/ip/xpu/src/spi.v}. The on/off trigger is given by the low MAC logic in FPGA.
\subsection{Turnaround time}
\begin{figure}
    \centering
    \includegraphics[width=0.75\linewidth]{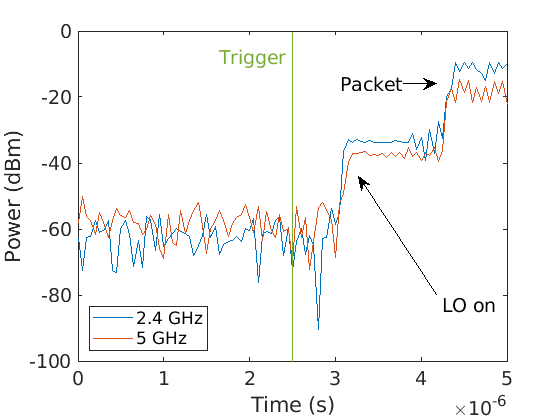}
    \caption{Tx power over time measured around Tx LO on command.}
    \label{fig:tx_lo_on}
\end{figure}

\begin{figure}
    \centering
    \includegraphics[width=0.77\linewidth]{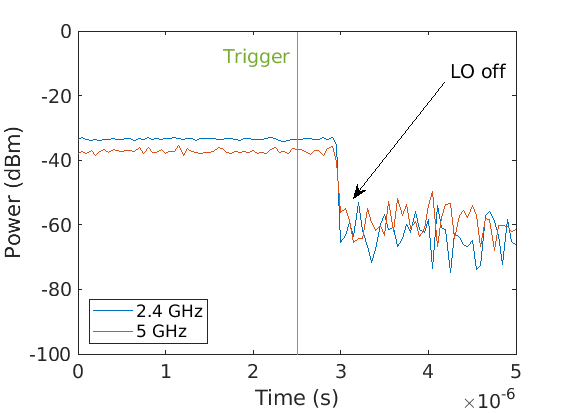}
    \caption{Tx power over time measured around Tx LO off command.}
    \label{fig:tx_lo_off}
\end{figure}

For quantifying TT, the Tx power has been measured using a signal analyzer (Anritsu MS2690A \cite{Anritsu}) with a \SI{10}{MHz} span and by setting the center frequency equal to the Tx LO frequency. A General Purpose Input/Output (GPIO) pin on the SoC is connected to the signal that initiates the SPI write. This pin is used as trigger for the signal analyzer, which then measures the Tx power \SI{2.5}{\micro \second} before and after the time at which the SPI write was initiated. Measurement points are taken every \SI{0.05}{\micro \second}. We have measured the Tx power for two LO frequency configurations:  Wi-Fi channel 1 (\SI{2.4}{GHz} band) and channel 44 (\SI{5}{GHz} band).
The output power when turning the Tx LO on and off is shown in Figure \ref{fig:tx_lo_on} and Figure \ref{fig:tx_lo_off} respectively.
In Figure \ref{fig:tx_lo_on}, it can be seen that after the trigger (green vertical line), the power shoots up and quickly stabilizes, indicating that the LO is turned on. The second jump indicates the start of a packet transmission. The time between the trigger and when the Tx LO is powered up is about \SI{0.65}{\micro \second}, while the time between the trigger and when the Tx LO is powered down is only about \SI{0.5}{\micro \second} as shown in Figure \ref{fig:tx_lo_off}. 

Thus, the calculated TT in Section \ref{sec:solution} from Rx to Tx falls within the measurement resolution, while the TT from Tx to Rx down is even lower, leading to TT values comparable to what can be achieved using protocol-specific chips. TT seems further to be independent on the frequency configuration. However, the difference between the average power with LO on or off is about \SI{30}{dB} in the \SI{2.4}{GHz} band, whereas it is about \SI{22}{dB} in the \SI{5}{GHz} band.

\subsection{Self-interference} 
We assess the influence of self-interference by recording the noise floor at the Rx antenna. The expected decrease in noise floor is smaller than the difference in Tx power as measured above, since the noise will not leak to the Rx completely.  

We investigate three scenarios, namely FDD mode, Standard ENSM TDD Mode (referred to as TDD) and our LO control. Using the FDD mode, the SIFS requirement is met, but no effort is done on mitigating self-interference. The Standard ENSM TDD Mode is also selected since it is the recommended way to implement Rx-Tx switching achieving no self-interference, despite the large TT.

We utilize the trigger-based I/Q capture feature of \textit{openwifi} to obtain the noise floor (see application note \cite{openwifi}). During the measurement, the application is set to trigger always. Then the I/Q samples from the Rx antenna are constantly captured via a side channel in the FPGA. This module transfers the samples via DMA to the Linux driver. Using a user space application, these samples can be exported and analyzed offline.

\begin{figure}
    \centering
    \includegraphics[width=0.8\linewidth]{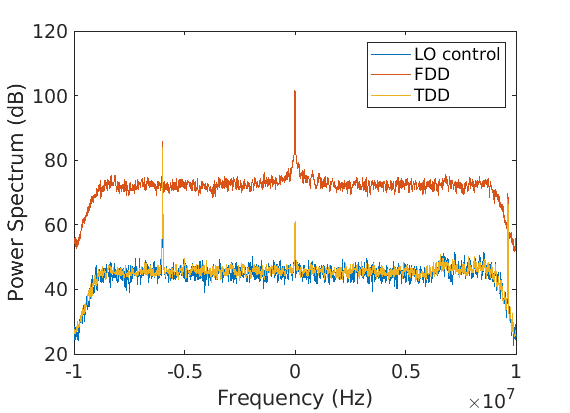}
    \caption{Receiver noise floor between different modes in the \SI{2.4}{GHz} band.}
    \label{fig:noise_2GHz}
\end{figure}

\begin{figure}
    \centering
    \includegraphics[width=0.8\linewidth]{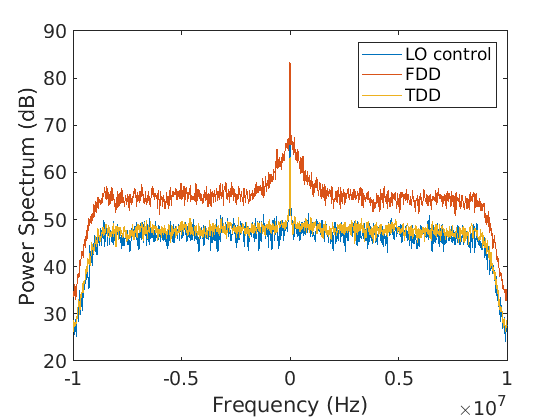}
    \caption{Receiver noise floor between different modes in the \SI{5}{GHz} band.}
    \label{fig:noise_5GHz}
\end{figure}

We utilized two VERT2450 antennas from Ettus Research \cite{Ettus}, connected to ports Tx1 and Rx1 of the AD9361, oriented orthogonal to each other. In each of the measurements, the LO leakage and quadrature calibration algorithms of the AD9361 are applied. The Automatic Gain Control (AGC) was set to a manual level of \SI{62}{dB} for all measurements. Unintentionally captured packets from the environment were filtered from the results. The noise floor for the different modes on channel 1 is shown in Figure \ref{fig:noise_2GHz}, while Figure \ref{fig:noise_5GHz} shows the noise on channel 44. 
The average relative power of the noise floor in the full \SI{20}{MHz} bandwidth for each mode is calculated and shown in Table \ref{tab:avgpow}.
\begin{table}[]
\vspace{-0.5cm}
\centering
\caption{Average relative power of the noise floor.}
\label{tab:avgpow}
\begin{tabular}{|l|ll|}
\hline
\multirow{2}{*}{\textbf{Method}} & \multicolumn{2}{l|}{\textbf{Average power (dB)}} \\ \cline{2-3} 
                        & \multicolumn{1}{l|}{2.4 GHz}    & 5 GHz   \\ \hline
FDD                     & \multicolumn{1}{l|}{66.4}     & 58.0        \\ \hline
TDD                     & \multicolumn{1}{l|}{53.3}     & 53.7        \\ \hline
LO control              & \multicolumn{1}{l|}{53.0}     & 53.4        \\ \hline
\end{tabular}%
\end{table}
It can be seen that for both frequency bands, the power of the noise floor for TDD is slightly higher than our LO control. When using FDD mode, the noise floor power is about \SI{13}{dB} higher in the \SI{2.4}{GHz} band and more than \SI{4}{dB} higher in the \SI{5}{GHz} band. The difference between the bands can be explained by the difference in Tx power as measured before. It is expected that the receiver sensitivity will be improved to the same extent. 

\section{Conclusion}
SDR platforms using ZIF transceivers suffer from self-interference in FDD mode using the same frequency for Rx and Tx. By turning the LO divider of the AD9361 on and off via an SPI writing operation in real-time, we have created a way to eliminate self-interference. This method decreases the receiver noise floor by about \SI{13}{dB} in the \SI{2.4}{GHz} band and more than \SI{4}{dB} in the \SI{5}{GHz} band as compared to FDD mode. This is a slightly better performance as compared to what can be achieved in conventional TDD mode. However, as apposed to this mode, which has a TT of at least \SI{55}{\micro \second}, our approach only takes \SI{0.64}{\micro \second} for Rx to Tx switching and just \SI{0.5}{\micro \second} for Tx to Rx switching. Our solution thus provides a way to implement TDD-based protocols on SDRs even with very strict latency requirements, while not having to compromise the receiver noise floor. For future work, we suggest to verify the impact on receiver sensitivity in over-the-air tests.

\section*{Acknowledgment}
This research was partially funded by the Flemish FWO SBO \#S003921N VERI-END.com project.

\bibliographystyle{IEEEtran}
\bibliography{references} 

\end{document}